# Topology-Induced Critical Current Enhancement in Josephson Networks

P.Silvestrini[1], R. Russo[1,2], V.Corato[1], B. Ruggiero[2,1], C.Granata[2], S. Rombetto[2,#], M.Russo[2], M. Cirillo[3], A. Trombettoni[4] and P.Sodano[4,5]

[1] *Dipartimento d'Ingegneria dell'Informazione, Seconda Università di Napoli, Aversa, Italy*

[2] *Istituto di Cibernetica "E. Caianiello" del CNR, Pozzuoli, Italy*

[3] *Dipartimento di Fisica and INFM, Università di Roma "Tor Vergata", 00173 Roma, Italy*

[4] *Dipartimento di Fisica, Università di Perugia, 06123 Perugia, Italy and Sezione I.N.F.N.-Perugia*

[5] *Progetto Lagrange, Fondazione C.R.T. e Fondazione I.S.I., Dipartimento di Fisica, Politecnico di Torino, Corso Duca degli Abruzzi 24, 10124 Torino, Italy*

Abstract:

We investigate the properties of Josephson junction networks with inhomogeneous architecture. The networks are shaped as "square comb" planar lattices on which Josephson junctions link superconducting islands arranged in the plane to generate the pertinent topology. Compared to the behavior of reference linear arrays, the temperature dependencies of the Josephson currents of the branches of the network exhibit relevant differences. The observed phenomena evidence new and surprising behavior of superconducting Josephson arrays as well as remarkable similarities with bosonic junction arrays.

Macroscopic Quantum Coherence occurs in a variety of physical systems such as Bose-Einstein condensates of atomic gases (1-3), superfluid Helium (4,5), superconductors (6,7), solid-state mesoscopic systems (8-12) and cold atoms in optical lattices (13-15). In this area, Josephson Junction Networks (JJN) are by now the prototype of a versatile solid-state system, which - by acting on a few control parameters - may be used for the engineering of a variety of macroscopically quantum states (16-19). Very often the JJN that have attracted attention for modelling of superconductive systems have been planar arrays obtained by closing superconducting loops by point Josephson junctions (16). This kind of arrays have been very helpful and provided much physical insight, for example, in the physics of granular superconductors; series arrays of junctions instead have provided important and useful devices in superconductive electronics (7) . In this paper we shall provide experimental evidence for the fact that *in* JJN with non-conventional architecture, containing no superconductive loops and made by different branches in which the junctions are series-connected, the network's topology (i.e. the way in which the superconducting islands of the array are coupled together by the Josephson junctions) plays a key role in determining new and unexpected properties of the array.

We have fabricated J*JN* with comb-like topology (see Fig.1) using a reliable niobium trilayer technology (20). The individual Josephson junctions forming the array have a 5x5 or 4x4 $\mu m^2$ area and critical current densities ranging from 30 to 50 A/cm$^2$, resulting in a relatively large capacitance (*C*~2pF) and Josephson critical current $I_c$~10μA (see Fig.2a). In this limit one can safely neglect the charging energy $E_c$ with respect to the Josephson energy $E_j$: $E_c = e^2/2C$, $E_j = \frac{\hbar}{2e} I_c$ and $E_c/E_j<0.001$ (being $e$ the electron charge and $\hbar$ the Plank constant). In Fig. 1b we show an optical image of a fabricated sample; in the photo we can see the square-comb network on which we have evidenced the localization of the Backbone Array (BBA) and the Central Finger Array (CFA). Beside the comb network, we have fabricated reference linear arrays called RBA (Reference Backbone Array) and RFA (Reference Finger Array) localized very close to the comb structure; the arrays are composed of superconducting islands connected to each other in series through Josephson junctions. The comparison between the critical currents in the reference arrays and the ones observed in the backbone and finger arrays of the square comb is a relevant tool for our investigation.

The Current-Voltage (I-V) characteristics were recorded by supplying a bias current to all the junctions of the arrays by means of contact pads located at the ends of the individual arrays (BBA, RFA etc.). For the data acquisition we used an experimental set-up filtered from the external noise, as tested in similar Josephson junction measurements.

The critical current of the reference arrays show a good uniformity as well as a modulation to zero in a magnetic field (see inset Fig.2b); these behaviors (7) represent an important indication in favor of the good quality of our fabrication process.

In Fig. 2 we show the I-V characteristics of the BBA (black line) of the network compared with the reference array, RBA (red line), at two different temperatures. Since the junctions are current-biased in series, the voltage readout across the whole array shall be the sum of the individual junction voltages. The switches of the first 10 junctions in series are enlarged in the inset of Fig.2a; the inset shows how the number of junctions in the array can be determined by counting the number of voltage jumps contributing to the gap voltage sum of the series connection. It is worth noticing that each voltage jump is roughly 2.7 mV, corresponding to the gap value of a single junction: we stress that the junctions which are not switched to the gap voltage are still in the superconducting (zero voltage) state. Fig. 2a shows that the Josephson critical current of the BBA is systematically higher than the reference one. At T=4.2K, the critical current difference between BBA and RBA is of the order of few percents, and, thus, it is not so significant; it could be ascribed to fabrication technology margins and related uncertainties (although we shall see in the following our uncertainties are lower, see figure 4).

In Fig. 2b we show, that, at a temperature of 1.2K, the critical current of each junction in the RBA array (red line in figure) increases as expected from the BCS-based Ambegaokar-Baratoff temperature dependence of the Josephson currents (7), whereas most of the junctions in the BBA show a critical current raising significantly above this expected BCS behavior. The two curves are almost identical from 0 to 50mV that is the sum of switches to the gap voltage of about 18 Junctions. There are18JJs in BBA that have the same critical current than 18JJs in the RBA. The remaining 54JJs of the BBA (that are still in the zero voltage state) have a critical current higher than the corresponding JJs in the RFA. Upon increasing the bias current also these junctions switch to the gap voltage manifesting a larger $I_c$ in comparison to the linear array. In particular after the switching from the superconducting state to the gap branch of approximately 32 JJs (corresponding to about 90mV), the difference in critical current between the remaining JJs in BBA and RBA is about 15%. There are, then, 40 JJs which are still in the superconducting state with a critical current about 15% higher than the ones in the reference array. We believe that some junctions in the BBA do not display the current rise most likely because our square combs have a finite size and the junctions with low critical currents are located at the edge of the comb.

From the data shown in Fig. 2b we can quantitatively estimate the critical current enhancement in the BBA introducing the parameter $F$ defined as $F = \frac{\langle I_c^{BBA} \rangle - \langle I_c^{RBA} \rangle}{\langle I_c^{RBA} \rangle}$, resulting in about 15% at T=1.2 K.

Here, $\langle I_c^{BBA} \rangle$ $(\langle I_c^{RBA} \rangle)$ is the average value of the critical current of BBA (RBA) and it is computed excluding the first 20JJs (the junction having a lower $I_c$, similar to the RFA due to the finite size of the square comb). This estimate, repeated for all our samples, provides an average result for the enhancement fraction $F$ of 13% with a 5% scatter around the average. Although the scatter around the average can be ascribed to uncertainties in fabrication and measurements, a difference of average critical current between the BBA and the RBA of 13% can only be attributed to *new effects induced by* the network topology.

In Fig.3 we display the I-V curves for the CFA of the network and its linear reference array RFA. In Fig. 3a we observe (similarly to what we see in Fig. 2a) that the Josephson current of the CFA does not differ significantly from the RFA at T= 4.2 K. In Fig. 3b, instead, we show that, at T=1.2K, the critical currents of junctions in the CFA undergo a non uniform variation. Moreover, we see that, at T=1.2K, most of the junctions in the CFA have critical current values lower than the RFA: thus, lowering the temperature, leads to opposite behaviors for the BBA and for the CFA. Most of the junctions in the CFA show an increase of the critical current smaller than the one expected from the Ambegaokar-Baratoff equation; in addition, a few junctions have a critical current at 1.2K lower than the lowest recorded at 4.2K. We note that, instead, the RFA follows the Josephson current increase predicted by the BCS theory *for a linear chain.* From the comparison between Fig. 2b and Fig. 3b we can say that the topology induces in the network a new temperature behavior, namely: *at low temperature,* $I_c$ *, increases along the backbone direction while it decreases along the fingers of the comb.* The data presented are well representative of a behavior that has been observed in all the ten samples which we measured and characterized.

From the reported evidences we have no doubt left that the observed effects are due to the peculiar topology of the network. However, as a further check of the reliability of our measurements we performed an additional experimental test. The goal of this test is to provide a measurement of the BBA and CFA when these are isolated from the remainder of the network; for this purpose, using lithographic techniques, we removed in three samples all the fingers from the comb network. Re-measuring the Josephson currents in these samples after the cutting process, we find again that the current of the RBA follows the BCS predictions at all temperatures (from 4.2K to 1.2K) but now the same happens the isolated BBA whose characteristic remains always identical to the one of the RBA (see Fig.4). Thus, the recorded differences of the Josephson critical current between the BBA and RBA disappear when the BBA is isolated from the surrounding network a result very well evidenced in Fig. 4. After the fingers removal, the critical current of both arrays (RBA and BBA) displays no significant differences both at 4.2K and 1.2 K: the different behaviors exhibited by the BBA and RBA shown in Fig. 2b can be solely attributed to the comb topology. We also applied the same lithographic procedure to the CFA obtaining similar results: measuring the I-V curves of the finger isolated from its surrounding we find that the temperature behavior of CFA is identical to that of the reference array.

Our experiments show remarkable similarities with behaviors expected to be observed for the systems of ultracold atoms, arranged in inhomogeneous comb-shaped optical lattices, analyzed in refs.18,19. An accurate quantitative explanation of the data presented in this paper should be sought within the framework of other theories/models specific of the superconducting state provided that one is able to tackle carefully the new effects induced by the existence of the hidden spectrum in a comb-graph.

We thank F. P. Mancini and M. Rasetti for stimulating and useful discussions, as well as E. Esposito for the help in setting the cryogenic facilities at the "Seconda Università di Napoli". Work partially supported by the M.I.U.R. National Project "JOSNET" (grant n.2004027555)

**REFERENCES**:

1. E. A. Donley, *et al.*, *Nature* **412**, 295 (2001).
2. J.R. Anglin, W. Ketterle, *Nature* **416**, 211 (2002).
3. F. Dalfovo, S. Giorgini, L. P. Pitaevskii, S. Stringari, *Rev. Mod. Phys.* **71**, 463 (1999).
4. A.J. Leggett, *Rev. Mod. Phys.* **71**, S318 (1999).
5. A.J. Leggett, *Rev. Mod. Phys.* **76**, 999 (2004).
6. J.R. Schrieffer, ( Wesley, Menlo Park, U.S.A., 1988); P.G. De Gennes, (Advanced Books Classics, Westview Press, U.S.A.,1999).
7. A. Barone, G. Paternò, *Physics and Applications of the Josephson Effect* (Wiley, New York, 1982); T. Van Duzer and C. W. Turner, *Principles of Superconductive Devices and Circuits* (Elsevier, NY 1998)

8. Chiorescu, Y. Nakamura, C.J.P.M. Harmans, J.E. Mooij, *Science* **299**, 1869 (2003).

9. D. Vion, *et al. Science* **296**, 886 (2002).

10. J. R. Friedman, V. Patel, W. Chen, S. K. Tolpygo, J. E. Lukens, *Nature* **406**, 43 (2000).

11. Y. Yu, S. Han, X. Chu, S.I. Chu, Z. Wang, *Science* **296**, 889 (2002).

12. J. M. Martinis, S. Nam, J. Aumentado, C. Urbina, *Phys. Rev. Lett* **89**, 117901 (2002).

13. D. Jaksch, C. Bruder, J. Cirac, C. W. Gardiner, P. Zoller, *Phys.Rev.Lett.* **81**, 3108 (1998).

14. F.S. Cataliotti, *et al. Science* **293**, 843 (2001).

15. M. Greiner, O. Mandel, T. Esslinger, T. W. Hansch, I. Bloch, *Nature* **415**, 39 (2002).

16. R. Fazio, H. van der Zant, *Phys. Rep.* **355**, 235 (2001).

17. Y. Makhlin, G. Schoen, A. Shnirman, *Rev. Mod. Phys.*, **73**, 357 (2001).

18. R. Burioni *et al.*, *Europhys. Lett.* **52**, 251-256 (2000); R. Burioni, D. Cassi, M. Rasetti, P. Sodano, A. Vezzani, *J. Phys. B* **34**, 4697 (2001).

19. G. Giusiano, F. P. Mancini, P. Sodano, A. Trombettoni, *Int. Jour. Mod. Phys. B* **18**, 691 (2004); I. Brunelli, G. Giusiano, F. P. Mancini, P. Sodano, A. Trombettoni, *J. Phys. B* **37,** S275 (2004).

20. C. Granata, *et al*. *Appl. Phys. Lett*. **80**, 2952 (2002). See also reference 9 cited therein.

21. B. Ruggiero, *et al.*, *Phys. Rev. B* **57**, 134 (1998).

**Figure 1** (Color online) a) Schematic drawing of a square comb array. The superconducting islands (full rectangles in the figure) are connected to each other in series through Josephson junctions JJ (crosses in the drawing). A central array named backbone array (BBA) is connected to linear arrays (fingers) perpendicular to it; there are no connections between fingers, except through the BBA. The black curve sketched in the figure represents the qualitative behavior of the ground eigenstate of the boson hopping Hamiltonian, having the maximum on the backbone(18,19). b) Optical microscope pictures of a fabricated sample. The Central Finger Array is referred as CFA. Linear arrays used as references (RBA and RFA) located very close to the junction network are also shown. In the inset the detail of a BBA region is well visible.

**Figure 2** (Color online) Current-voltage characteristics for the BBA (black curve) and RBA (red curve) including JJ with area A=5x5$\mu m^2$ a) measured at T= 4.2 K and b) measured at T=1.2 K. The x-axis represents the measured voltage of the series array of 72 junctions. The inset of Fig.2a shows a magnification of the critical current switching of the first 10 junctions in the 5x5$\mu m^2$ array whereas the inset in Fig. 2b reports the magnetic pattern of the two arrays (circles and squares) showing a modulation to zero for both arrays (black line is the theoretical sinx/x behavior)

**Figure 3** (Color online) Current-voltage characteristics for the Central Finger array, CFA (black curve), and the reference arrays, RFA (red curve), with junction area A= 5x5$\mu m^2$ a) measured at T=4.2 K and b) measured at T=1.2 K. The insets show a magnification of the critical current switching of the first 10 junctions in the 5x5$\mu m^2$ CFA at the T=4.2K and T=1.2K. Note the at 1.2K the lowest $I_c$s of CFA are lower than the lowest $I_c$s recorded at 4.2K on the same array.

**Figure 4** (Color online) Current-voltage characteristics of the same chip shown in Fig. 2 after removing the fingers from the comb network. The black curves represent the BBA after it has been isolated from the remainder of the network, and the red curve represents the RBA response during the same measurement run. a) I-V curves measured at T= 4.2K and b) at T=1.2K.

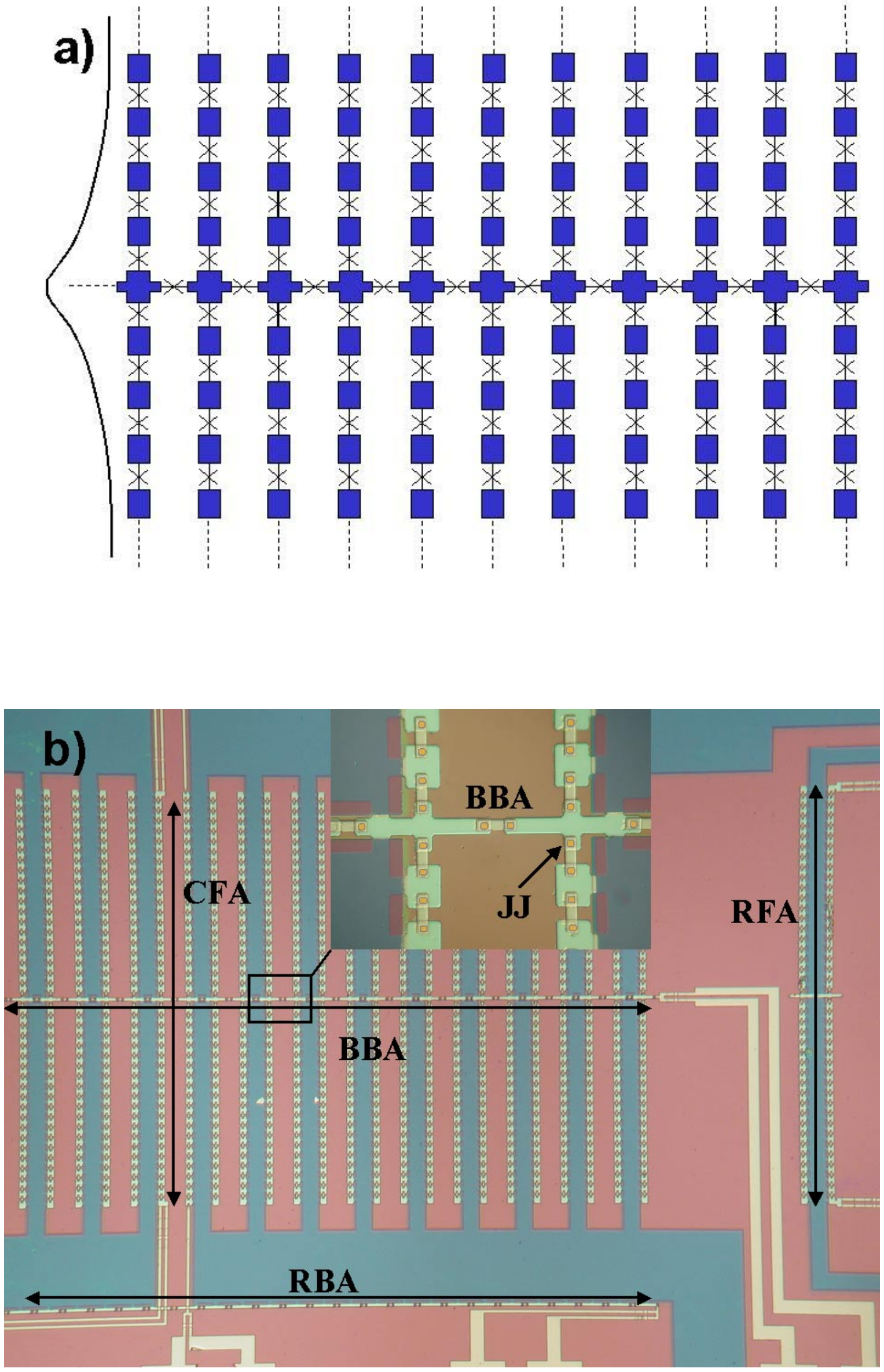

FIGURE 1

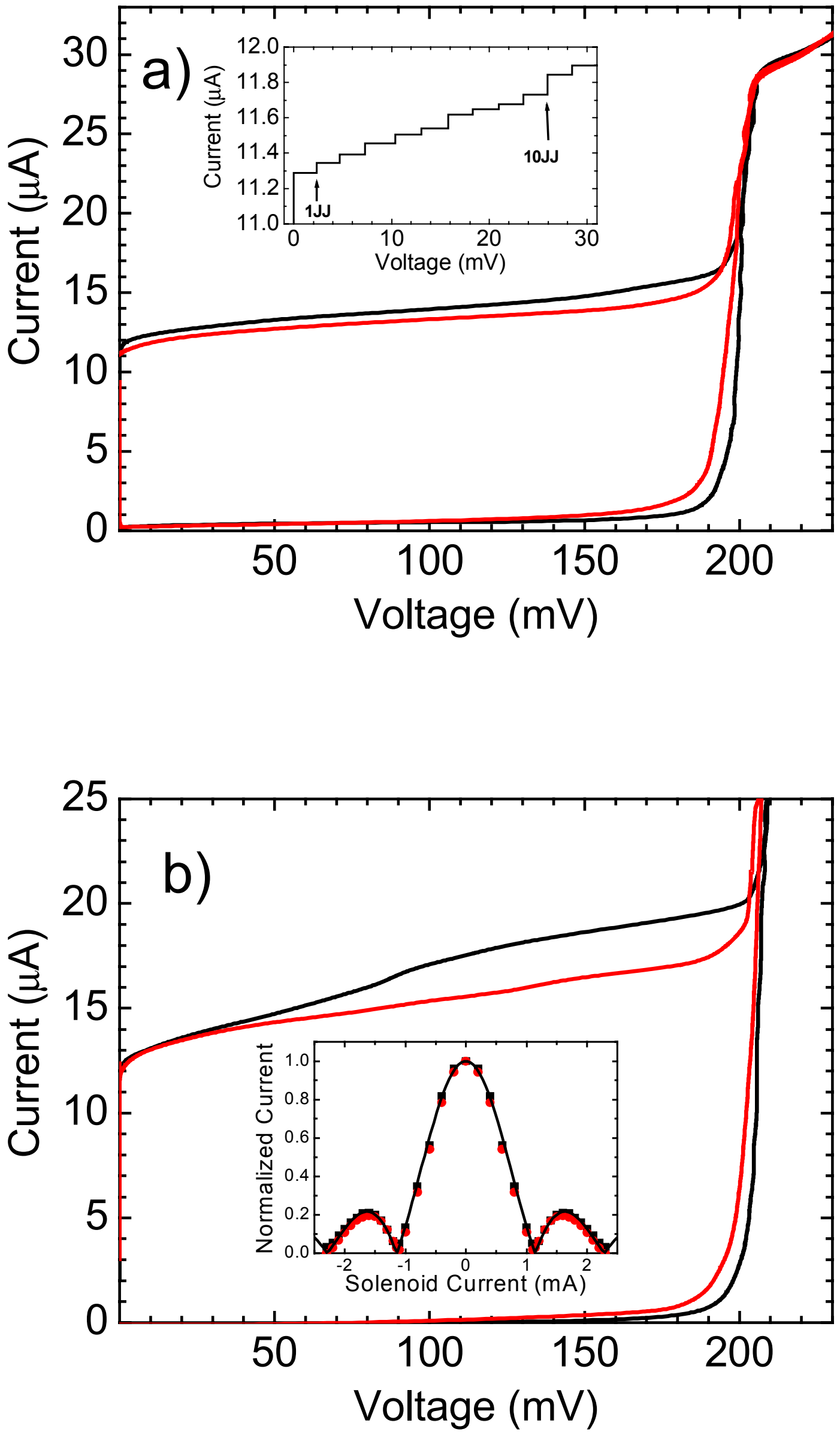

FIGURE 2

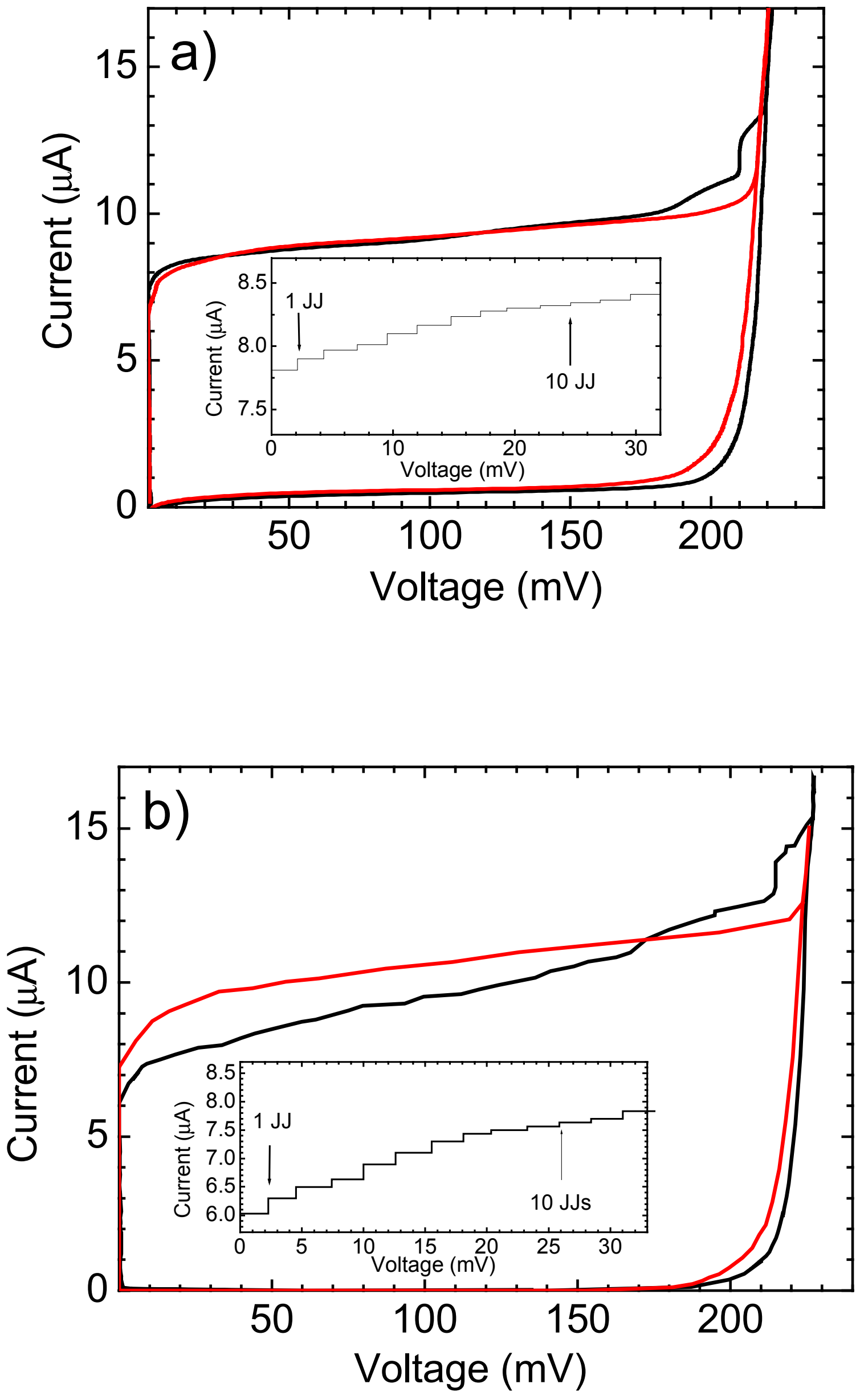

FIGURE 3

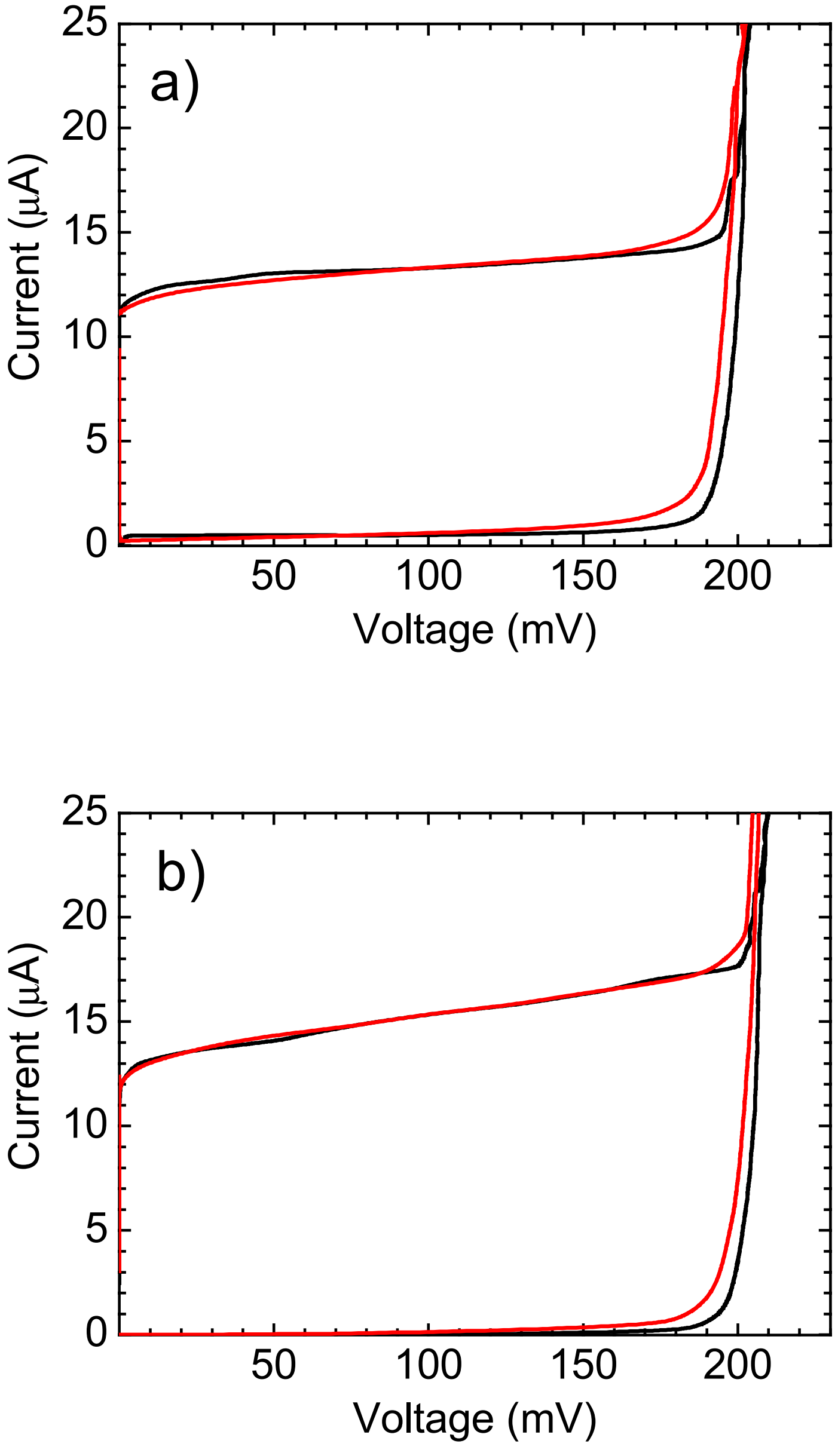

FIGURE 4